\documentclass[preprint]{aastex}
\usepackage{epsfig}

\newcommand{\beq}{\begin{equation}}
\newcommand{\eeq}{\end{equation}}
\newcommand{\beqa}{\begin{eqnarray}}
\newcommand{\eeqa}{\end{eqnarray}}
\newcommand{\msun}{\hbox {$M_{\sun}$ }}

\newcommand{\dv}{\hbox {$\Delta \rm V^{TO}_{HB}$}}
\newcommand{\dvbto}{\hbox {$\Delta \rm V^{SGB}_{HB}$}}

\newcommand{\mvrr}{\hbox {$\rm M_v(RR)$}}

\newcommand{\rgc}{\hbox {R$_{\rm GC}$}}
\newcommand{\ea}{{\it et al. }}
\newcommand{\feh}{\hbox{$ [{\rm Fe}/{\rm H}]$}}
\newcommand{\ebv}{\hbox {${\rm E(\bv)}$}}
\newcommand{\dmv}{\hbox{$(m\!-\!M)_{V}$}}

\newcommand{\vi}{\hbox{$V\!-\!I$}}

\newcommand{\dyz}{\hbox {$\Delta Y/\Delta Z$}}
\newcommand{\evi}{\hbox {${\rm E(\vi)}$}}

\begin{document}

\title{The Age of the Inner Halo Globular Cluster NGC 6652\altaffilmark{1}}

\altaffiltext{1}{Based on observations made with the NASA/ESA Hubble
Space Telescope obtained at the Space Telescope Science Institute,
which is operated by the Association of Universities for Research in
Astronomy, Incorporated, under NASA contract NAS5-26555.}

\author{Brian Chaboyer}
\affil{Department of Physics and Astronomy, Dartmouth College,
6127 Wilder Laboratory, Hanover, NH, USA 03755-3528}
\email{Brian.Chaboyer@Dartmouth.edu}

%\and

\author{Ata Sarajedini\altaffilmark{2}}
\affil{Astronomy Department, Wesleyan University, Middletown, CT 06459}
\email{ata@astro.wesleyan.edu}
\altaffiltext{2}{Guest User, Canadian Astronomy Data Centre, which is
operated by the Dominion Astrophysical Observatory for the National
Research Council of Canada's Herzberg Institute of Astrophysics.}

\and 

\author{Taft E. Armandroff} 
\affil{National Optical Astronomy
Observatory, P.O. Box 26732, Tucson, AZ 85726}
\email{armand@noao.edu}

\begin{abstract}
HST ($V,\, I$) photometry has been obtained for the inner halo
globular cluster NGC 6652.  The photometry reaches $\sim 4$ mag below
the turn-off and includes a well populated horizontal branch (HB).
This cluster is located close to the Galactic center at $\rgc \simeq
2.0\,$kpc with a reddening of $\evi = 0.15\pm 0.02$ and has a
metallicity of $\feh \simeq -0.85$.  Based upon \dvbto, NGC 6652 is
$11.7\pm 1.6$ Gyr old.  Using \dvbto\, precise differential ages for
47 Tuc (a thick disk globular), M107 and NGC 1851 (both halo clusters)
were obtained. NGC 6652 appears to be the same age as 47 Tuc and NGC
1851 (within $\pm 1.2$ Gyr), while there is a slight suggestion that
M107 is older than NGC 6652 by $2.3\pm 1.5\,$Gyr.  As this is a less
than $2\,\sigma$ result, this issue needs to be investigated further
before a definitive statement regarding the relative age of M107 and
NGC 6652 may be made.  

\end{abstract}


\section{Introduction}
NGC 6652 is a compact globular cluster which is projected near the
Galactic center at $\ell = 1.53\degr$ and $b = -11.38\degr$
\citep{har96}.  A color magnitude diagram (CMD) for this cluster was presented
by \citet[hereafter OBB]{ort94}, who determined a number of cluster
parameters, including $\feh \approx -0.9$, $\ebv = 0.10\pm 0.02$,
$<{\rm V_{HB}}> = 15.85\pm 0.04$ and $\dv = 3.35\pm 0.16$.  The small
value of \dv\ suggested that NGC 6652 is younger than the average
Galactic halo cluster. Adopting an absolute horizontal branch (HB)
magnitude  of ${\rm M_V^{HB}} = +0.7$, OBB determined $\rgc =
2.1\,$kpc.  Its proximity to the Galactic center, reasonably low
metallicity and small reddening make NGC 6652 somewhat unique.  It is an
inner halo globular cluster for which one can obtain an accurate
relative age with respect to other halo clusters.

Based upon its metallicity, radial velocity, HB type and position in
the Galaxy, \citet{zinn93} classified NGC 6652 as an old halo cluster.
This is at odds with OBB's suggestion (based on the turn-off
magnitude) that NGC 6652 is younger than the average halo cluster.
Because NGC 6652 is a compact cluster located in a crowded field, 
the photometry near the turn-off obtained by OBB had a great
deal of scatter leading to a large error in the age
determination. For this reason, we were granted HST time to obtain a deep
CMD of NGC 6652, in order to clearly delineate the turn-off region
and to obtain an accurate estimate of its age.  The observations and
data reduction procedure are presented in \S \ref{obs} and the
generation of the isochrones used to determine the age of NGC 6652 is
discussed in \S \ref{theory}.  The analysis of the CMD is presented in
\S \ref{cmd}, the discussion of the age of NGC 6652 is in \S \ref{age},
and the summary of the results are in \S \ref{summ}.

\section{Observations and Data Reduction \label{obs}}
\begin{figure}
\centerline{\epsfig{file=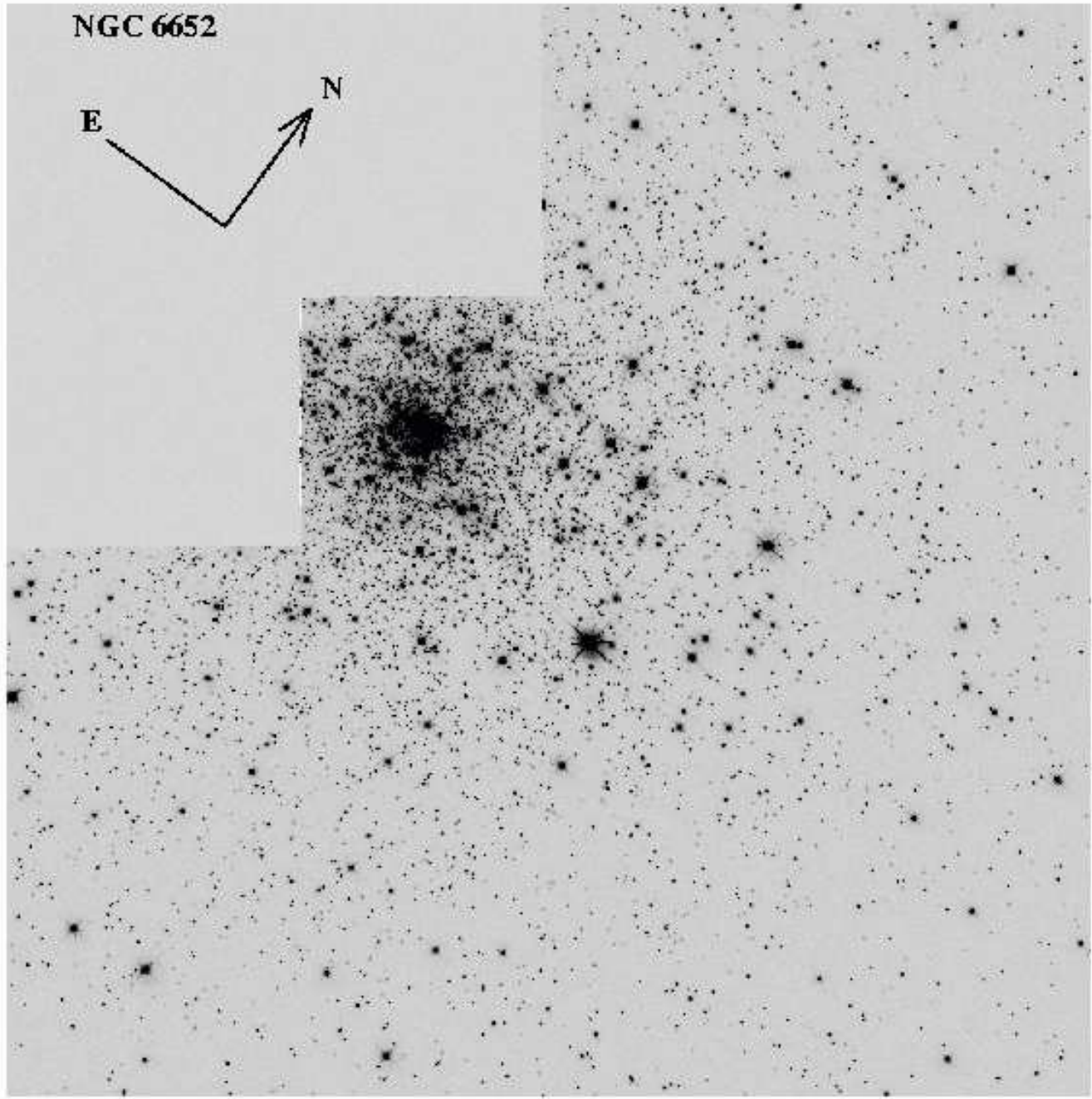,height=19.0cm}}
\caption{The averaged long exposure $V$ frame of NGC 6652. }
\label{figpict}
\end{figure}

NGC 6652 was observed in September 1997 with WFPC2 on HST.  The
cluster was centered on the PC1 chip.  Observations were obtained with
the F555W (V) filter ($3\times 23$ s and $12\times 160$ s) and the
F814W (I) filter ($3\times 20$ s and $12\times 160$ s).  The short
exposure times were chosen to ensure good photometry at the level of
the HB, while the long exposure times were picked to ensure good
photometry around the turn-off and to allow for a good
inter-comparison between the long and short exposure frames.  Figure
\ref{figpict} shows the averaged, long exposure F555W image of the cluster.
The WFPC2 images of NGC 6652 were divided into three sets of four long
exposure (160s) F555W and F814W frames and one set of three short
exposure F555W (23s) and F814W (20s) frames.  Each set of four long
and three short exposures in a given filter was averaged using
GCOMBINE in IRAF/STSDAS with cosmic ray rejection enabled and the data
quality files used to flag defective pixels. This yielded a total of 8
WFPC2 images which were input into our photometric procedure.

We utilized the aperture photometry routines in DAOPHOT II
\citep{stet1,stet2} adopting an aperture radius of 2.5 pixels in the
Planetary Camera and 2 pixels in the Wide Field chips. Corrections
from the small aperture photometry to the standard 0.5 arcsec radius
were determined using 30 to 50 relatively uncrowded stars. These gave
aperture corrections with typical standard deviations between 0.01 and
0.03 mag. The V and I instrumental magnitudes were matched to form
colors, and corrected to a 0.5 arcsec radius aperture. The total
instrumental magnitudes were then adjusted for exposure time and
placed on the standard VI system using the coefficients in Table 7
of \citet{holtz}. A correction was also applied for the well-known
charge transfer efficiency (CTE) problem using the formulation of
\citet{whit}. Magnitudes of stars in common between the long exposure
frames were averaged together and only stars measured on all three
frame pairs were retained. The final photometry file contains stars
fainter than V=17 from the long exposure frames and brighter than V=17
from the short exposures (4832 stars in total).  In order to ensure
that the long and short exposures magnitudes were internally
consistent, the magnitude of a number of stars (approximately 10 per
chip) in common between the long and short exposures were compared and
the average offset determined.  This average offset (0.03 mag) was
then applied to the short exposure magnitudes, implying that all of
the magnitudes are on the long exposure scale.  The photometry data
is presented in Table \ref{tabphot}, the full version of which is
available in the electronic version of this article.  

Due to its compact nature, the inner regions of NGC 6652 are 
crowded on the PC1 chip.  The faint (long exposure) photometry within
$r < 11.5''$ (250 pixels) of the cluster center showed considerable
scatter around the turn-off region.  For this reason, the long
exposure photometry in our final CMD only includes stars 
with $r > 11.5''$ from the cluster center.  This trimmed data-set
includes 3790 stars with good photometry. 

A careful inspection of the photometry obtained with the different
chips showed no difference [$\delta(\vi) < 0.005$] in the location of the
principal features (i.e.\ location of the main sequence, sub-giant
branch, red giant branch and the HB) in the CMD.  This indicates that
the aperture corrections (applied to each chip individually) are
accurate, implying that the internal errors in the photometry are very
small.  The external errors are set by the accuracy of the Holtzman
\ea (1995) transformations to the standard system.

%differences in the principal location various features in in the CMD
%between the different chips.  For example, the estimated color of the
%turn-off varied from $(\vi)_{\rm TO} = 0.800$ (PC1) to $(\vi)_{\rm TO}
%= 0.815$ (WF2 and WF4) to $(\vi)_{\rm TO} = 0.829$ (WF4).  The
%estimated V magnitude of the point $0.05$ redder than the turn-off and
%brighter varied from ${\rm V(BTO)} = 19.11$ (PC1, WF2) to  ${\rm
%V(BTO)} = 19.12$ (WF3) to  ${\rm V(BTO)} = 19.13$ (WF3). A check of
%the photometry indicated that these small differences were not
%correlated with distance from the cluster center, and so are must
%likely due to small errors  in the apertures corrections between the
%various chips.  This comparison indicates, that aside from the
%calibration uncertainties, the photometry likely has an overall
%uncertainty of $\pm 0.01$ mag in $V$ and $\pm 0.015$ mag in \vi\ due
%to uncertainties in the aperture corrections.  .  

\section{The Stellar Models and Isochrones \label{theory}}
Stellar evolution tracks for masses in the range $0.50 - 1.1 \,\msun$
(in $0.05\,\msun$\,increments) were calculated using the Yale stellar
evolution code \citep{yrec}.  These models incorporate the following
physics: high temperature opacities from \citet{opal}; low temperature
opacities from \citet{kurmol}; nuclear reaction rates from
\citet{bah92} and \citet{bah89}; helium diffusion coefficients from
\citet{mp93}; and an equation of state which includes the
Debye-H\"{u}ckel correction \citep{yrec}. Note that this equation of
state yields stellar models which are in good agreement with those
derived using the OPAL equation of state \citep{rog94,cha95}.  The
surface boundary conditions were calculated using a grey $T-\tau$
relation.  The
stellar models employ a solar calibrated mixing length ($\alpha_\sun =
1.78$). The models were typically evolved from the zero age main
sequence to the upper giant branch (around $M_V \approx -1.0$) 
in 4000 time steps.  In each time
step, the stellar evolution equations were solved with a numerical
accuracy exceeding 0.01\%.  The models did not include any
overshooting beyond the formal edge of the convection zones.

The heavy element composition of the models was chosen to reflect the
observed abundance ratios in metal-poor stars \citep[e.g.,][]{lam89}.
In particular, the $\alpha$-capture elements (O, Mg, Si, S, and Ca)
and Ti were enhanced by $+0.4\,$dex, while Mn was made to be deficient
by the same proportions.  The solar abundances (before
$\alpha$-enhancement) were taken from \citet{gre93}.  Kurucz (low
temperature) and Iglesias \& Rogers (high temperature) provided us
with opacities for this specific mixture.  

NGC 6652 has $\feh = -0.96$ on the \citet{zinnwest} scale and $\feh =
-0.85$ on the \citet{gratton} scale.  Based upon the inclination of
the red giant branch, OBB found that NGC 6652 was more metal-poor than
47 Tuc, and concluded that $\feh \approx -0.9$.  In light of these
abundance determinations, stellar models and isochrones were
calculated for $\feh = -1.20,\; -1.00,\; -0.85,\; -0.70$.  The most
metal-poor isochrone was calculated to facilitate an age comparison
with somewhat more metal-poor halo clusters (see \S \ref{btoage}).  To
explore the effect of the $\alpha$-element enhancement, at $\feh =
-0.85$, scaled solar abundance models and isochrones were calculated
in addition to the $\alpha$-element enhanced isochrones.

Our calibrated solar model had an initial solar helium abundance of $Y_\sun
= 0.263$ and heavy element mass fraction of $Z_\sun = 0.0182$.  Assuming a
primordial helium abundance of $Y_p = 0.234$ \citep{olive97}, our
calibrated solar model implies $\dyz = 1.59$.  Thus, the helium
abundance for the models was determined using
\begin{equation}
Y = 0.234 + 1.59\cdot Z
\label{yeqn}
\end{equation}
There is some evidence that 47 Tuc has a solar helium abundance
\citep{sw98}.  As 47 Tuc is being used as a comparison cluster in this
study, for $\feh = -0.70$ stellar models and isochrones were
calculated for a solar helium abundance ($Y_\sun = 0.263$) in addition
to the helium abundance implied by equation (\ref{yeqn}), $Y = 0.245$.
The evidence for a solar helium abundance in 47 Tuc is not strong, and indeed 
\citet{dorman} found $Y \approx 0.24$ from fitting the HB morphology.

The isochrones were constructed by interpolating among the
evolutionary tracks using the method of equal evolutionary points
\citep{prather} for the age range 6 -- 18 Gyr, in 1 Gyr increments.
The isochrones were transformed from the theoretical $(\log L,\, \log
{\rm T_{eff}})$ plane to observed colors and magnitudes using color
transformations and bolometric corrections based on the \citet{kuratm}
model atmospheres.  The transformations used here were kindly supplied
to us by Sukyoung Yi (private communication), who derived them for a
wide range of metallicities, using a procedure similar to that
described by \citet{bes98}.  At solar metallicities, the Yi tables
agree with \citet{bes98} Table 1 to within $\sim$0.007 mag in \bv\ and
$\sim$0.004 in \vi\ over the range of colors appropriate for old
clusters.  In the Yi tables, the solar colors are  
$(\bv)_{\odot} = 0.670$ and $(\vi)_{\odot} = 0.718$.
\citet{bes98} extensively tested their Kurucz-based
(ATLAS9) colors for solar metallicities, and found that all indices
agreed extremely well with observations for $\log {\rm T_{eff}} >
4250$K  (4250K corresponds to $\vi \simeq 1.3$ at $\feh = -0.85$).

\citet{ws99} recently studied the issue of color transformations for
isochrones in the VI-plane.  They concluded that on the main sequence
the color transformations based on the ATLAS9 models were a good choice.
On the giant branch \citet{ws99} preferred to use empirical color
relations.  However, an inspection of their Figure 5 indicates that
the color transformation based upon the ATLAS9 models is in good
agreement with the empirical relations preferred by \citet{ws99}. Thus
it appears that our choice of the Yi color table is a reasonable
one, in good agreement with presently available data. 

\begin{figure}
\centerline{\epsfig{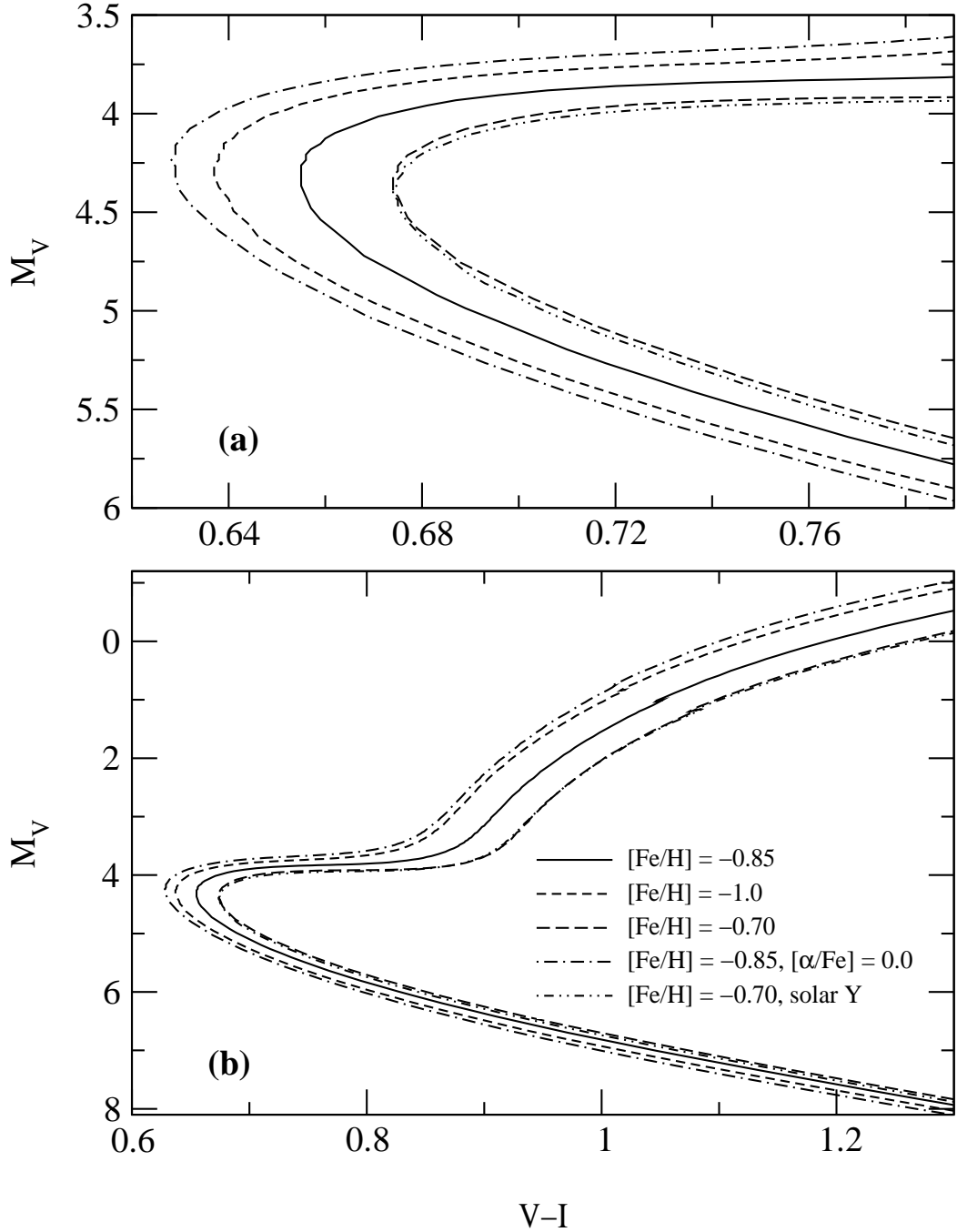}}
\caption{The ($M_V,\vi$) CMD of the 12 Gyr theoretical isochrones used
in this paper.  Unless otherwise noted in the legend, the isochrones
have $[\alpha/{\rm Fe}] = +0.4$ and a helium mass fraction $Y$ given
by equation (\ref{yeqn}).  The lower panel (b) shows the isochrones
over the entire range of validity of the color transformation ($\vi
\le 1.3$), while the upper panel (a) is an expanded view near the
turn-off.  }
\label{figiso}
\end{figure}

Figure \ref{figiso} displays the 12 Gyr isochrones for the various
compositions used in this paper.  It is interesting to note that
changing the helium abundance from $Y = 0.245$ (given by eqn.\
(\ref{yeqn})) to the solar value ($Y_\sun = 0.263$), has very little
effect on the isochrone.  Thus, the exact choice of $Y$ will not
effect the derived ages.  In contrast, changing the $[\alpha/{\rm
Fe}]$ ratio by $0.4\,$dex has a substantial effect on the isochrones
(larger than changing \feh\ by $0.15\,$dex) and so the uncertainty in
the actual value of $[\alpha/{\rm Fe}]$ in NGC 6652 will lead to a
substantial uncertainty in the derived age.

\begin{figure}
\centerline{\epsfig{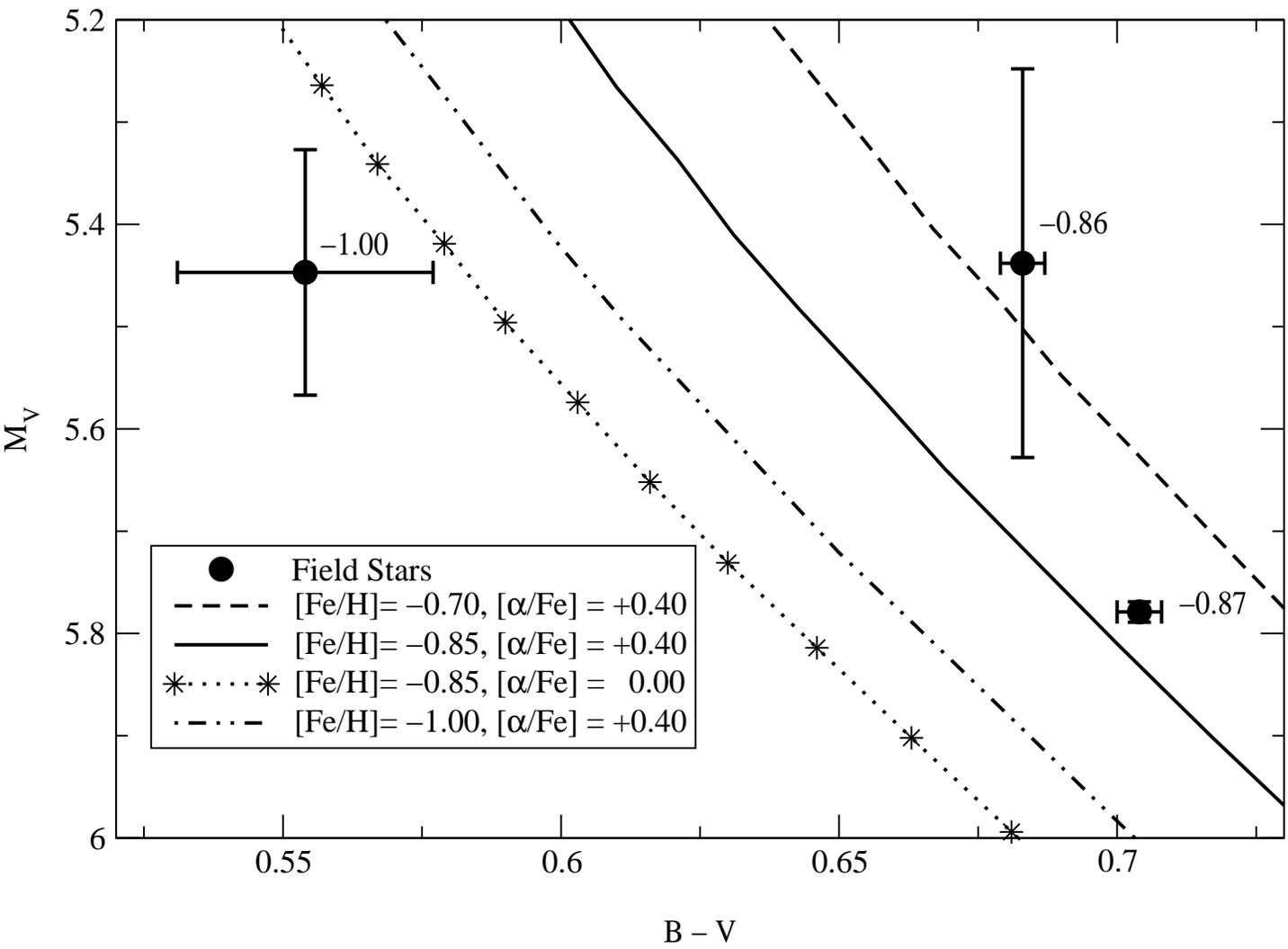}}
\caption{A comparison between unevolved, single stars in the Hipparcos
catalogue (with  $-1.05 \le \feh \le -0.65$ and good parallaxes, 
$\sigma_\pi/\pi < 0.10$)  with the isochrones presented in this
paper.  Each of the stars (HD  6582, 193901, 216179) has been labeled
with its \feh\ value determined from high dispersion spectroscopy.}
\label{fighipp}
\end{figure}

In \S \ref{isofit}, the isochrones are fit to the CMD of NGC 6652.  
This fitting procedure assumes that the location of the unevolved main
sequence predicted by the theoretical isochrones is accurate. To test this
assumption, the Hipparcos catalogue was searched for stars which (a) have
$\sigma_\pi/\pi < 0.10$, (b) are fainter than ${\rm M_V} \simeq 5.5$, and
(c) are not known or suspected Hipparcos binaries or variables.  This
resulted in a list of 2618 stars, of which the great majority have near
solar metallicity.  As we are interested in stars with a similar
metallicity to NGC 6652, we require stars with $-1.05 \le \feh \la -0.65$.  
To identify these metal-poor stars in the Hipparcos sample, we have cross
identified the above Hipparcos subsample with the 1996 high resolution
spectroscopic catalogue of \cite{cayrel}, and more recent papers which give
high dispersion abundance analysis of metal-poor stars: \cite{reid,
clemtini, tomkin, carretta}. Any stars which were known or suspected
binaries and whose companions would contribute a significant flux in the
BVI passbands were removed from our final list.  In total only 3 stars in
the Hipparcos catalogue pass our stringent selection criteria.  A
comparison between these stars and our isochrones is shown in Figure
\ref{fighipp}.  Unfortunately, we were unable to locate 
Kron-Cousins \vi\ colors for these stars, so the comparison is only
shown in \bv .  
The small number of stars prevents us from reaching any definitive
conclusions based upon this comparison.  However, we note that the star
with the most accurate parallax (with $\feh = -0.87$) lies close to our 
$\feh = -0.85$ isochrone.  Given the typical uncertainties in the
abundance determinations, $\sigma_{\feh} \approx 0.08\,$dex, it appears
that our isochrones correctly predict the location of the unevolved main
sequence around $\feh = -0.85$.

Further support that our isochrones correctly predict the location of the
unevolved main sequence comes from the following additional tests: (1) a
comparison between single field stars with good parallaxes from Hipparcos
and our isochrones found good agreement between $-1.5 \le \feh \le -1.0$
\citep{chahipp}. This study was interested in metal-poor globular
clusters, and so did not consider stars with $\feh > -1.0$.  (2) With no
adjustable parameters, our isochrones provide a good match to the main
sequence of the Hyades ($\feh = +0.15$) derived from Hipparcos parallaxes
\citep{cha6791}, and (3) we are using a solar calibrated mixing length
which implies our models are in good agreement with main sequence solar
metallicity stars.  Clearly, these tests are not definitive as the number
of single, metal-poor stars with well determined parallaxes is relatively
small.  However, these tests do show that the isochrones are not in gross
disagreement with the observations, and suggest that at the level of 
$\approx \pm 0.10$, the isochrones correctly predict the location of the
unevolved main sequence in relatively metal-poor stars ($-1.5 \le \feh \le
0.65$).

\section{Analysis of the CMD \label{cmd}}
\begin{figure}
\centerline{\epsfig{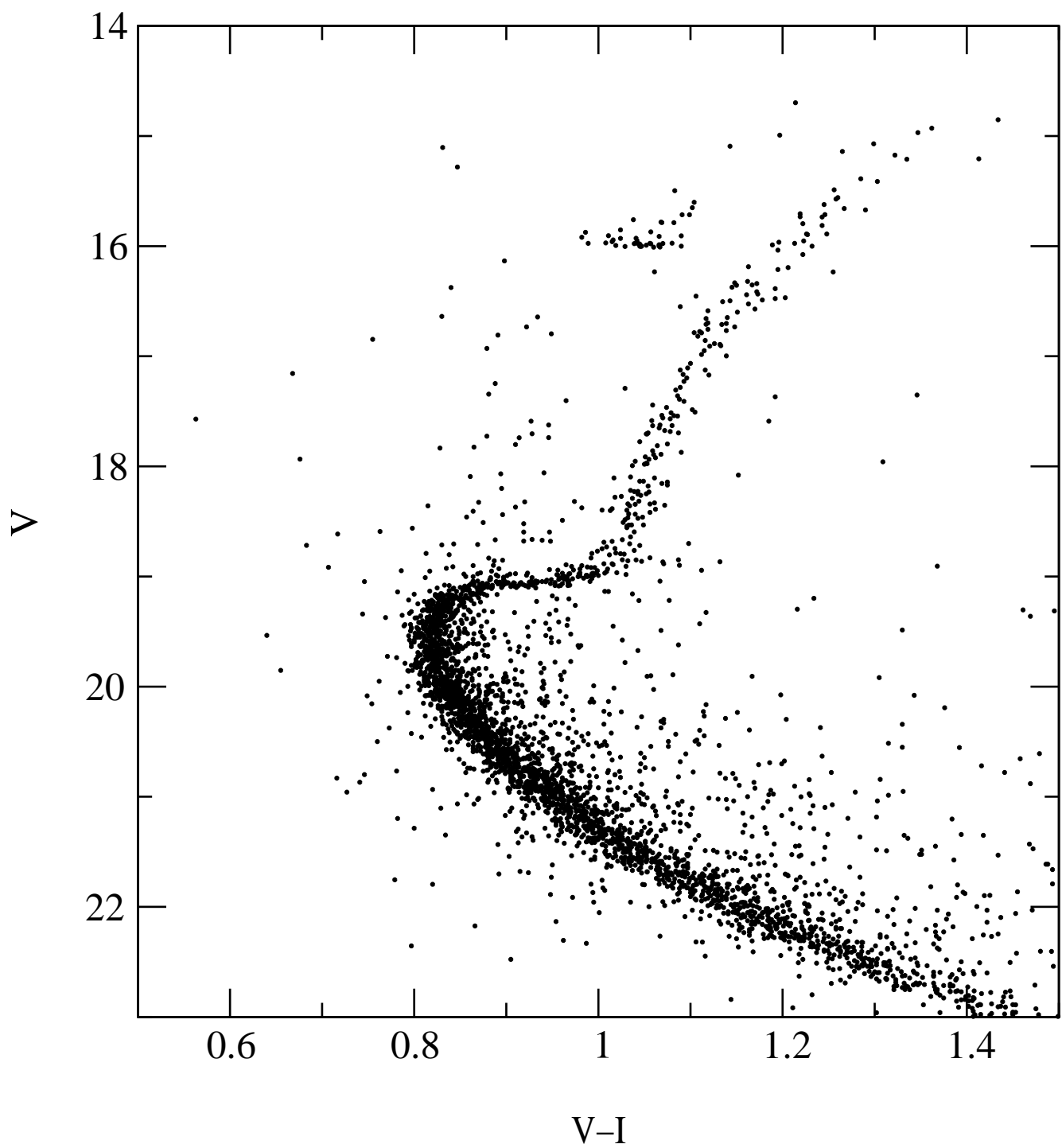}}
\caption{The ($V,\vi$) CMD  of NGC 6652.  }
\label{figcmd}
\end{figure}

The ($V,\vi$) CMD for NGC 6652 obtained from the HST data is shown in
Figure \ref{figcmd}.  As discussed in \S \ref{obs}, data from the long
exposures ($V \ge 17$) does not include stars with $r < 11.5''$. The CMD
clearly delineates the major evolutionary sequences from $\sim 4$ mag
below the turn-off to $\sim 1.5$ mag above the HB.  An inspection of
the CMD leads to the following values: $V_{\rm ZAHB} = 16.00\pm 0.03$,
$<V_{\rm HB}> = 15.96\pm 0.04$, $(\vi)_{\rm TO} = 0.818\pm 0.004$,
$V_{\rm TO} = 19.55\pm 0.07$,  $V_{\rm SGB} = 19.09\pm 0.03$,
$\dv =  3.59\pm 0.08$, and $\dvbto = 3.13\pm 0.05$.  All
errors are estimated internal errors, which reflect the scatter of the
photometry.  The magnitude of the HB was estimated at the mid-point in
the color of the HB (around $\vi \simeq 1.05$).  
The point $V_{\rm SGB}$ was defined by \citet{cha96} to
be the magnitude of the point on the sub-giant branch which is 
0.05 mag redder than the turn-off.  As discussed by \citet{cha96} it
is an excellent age indicator for old stellar systems.  Recently, 
\citet{ferr99} have presented an homogeneous catalog of HB parameters.
They stress that due to the rapid evolution away from the zero-age
horizontal branch (ZAHB), the observed lower envelope of the HB does
not coincide with the ZAHB from theoretical models.  The $V_{\rm
ZAHB}$ level reported above is the lower envelope of the
observations.  Using their theoretical HB models, \citet{ferr99}
determined that a correction of $\simeq +0.04$ mag needs to be applied to
the observed lower envelope of the NGC 6652 HB.  If such a correction
is necessary in our dataset then $V_{\rm ZAHB} = 16.04\pm 0.03$ using
the methodology of \citet{ferr99}.  

\begin{figure}
\centerline{\epsfig{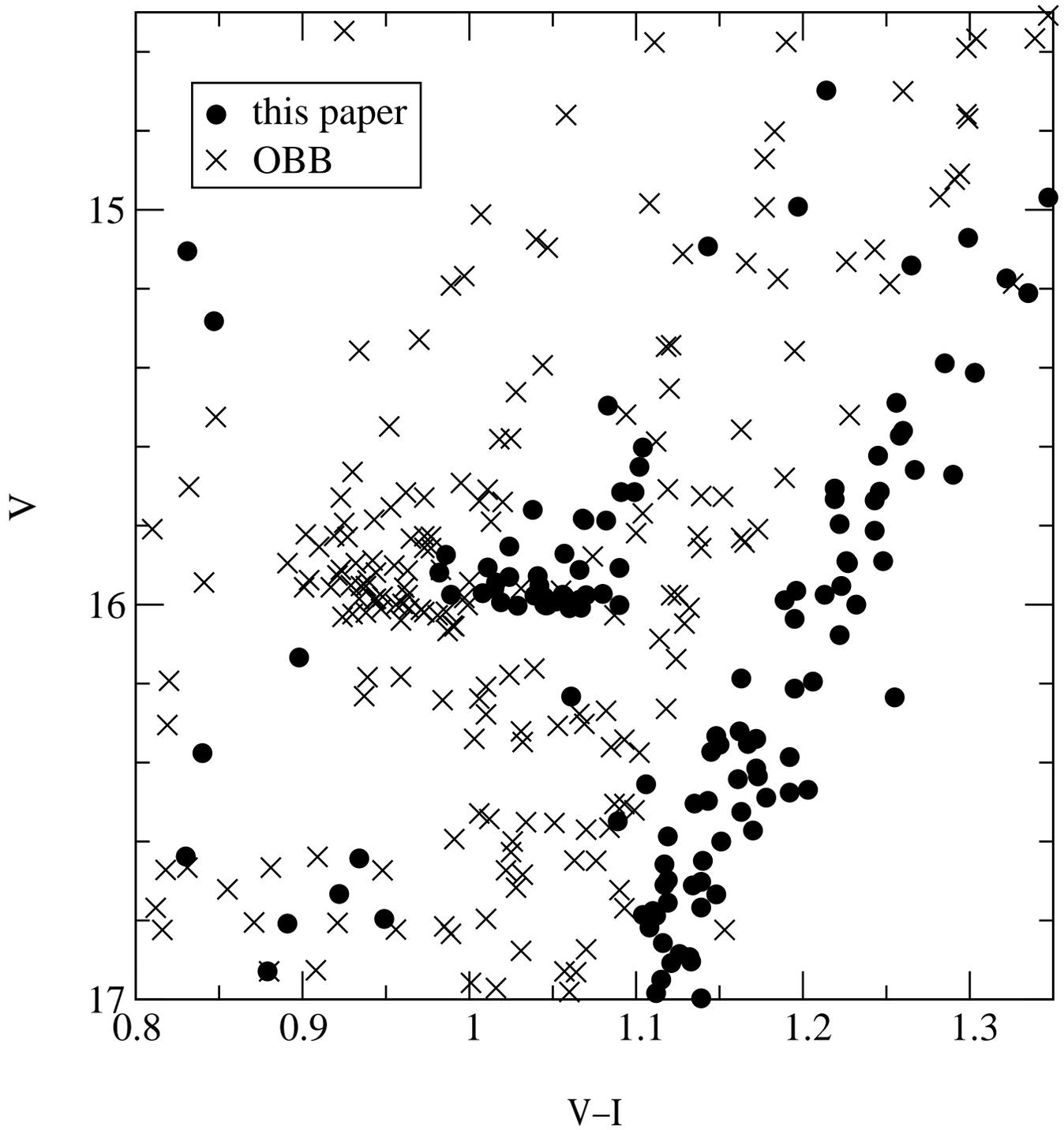}}
\caption{A comparison of the photometry near the HB presented in this
paper, to that presented by OBB.  }
\label{figort}
\end{figure}

\begin{figure}
\centerline{\epsfig{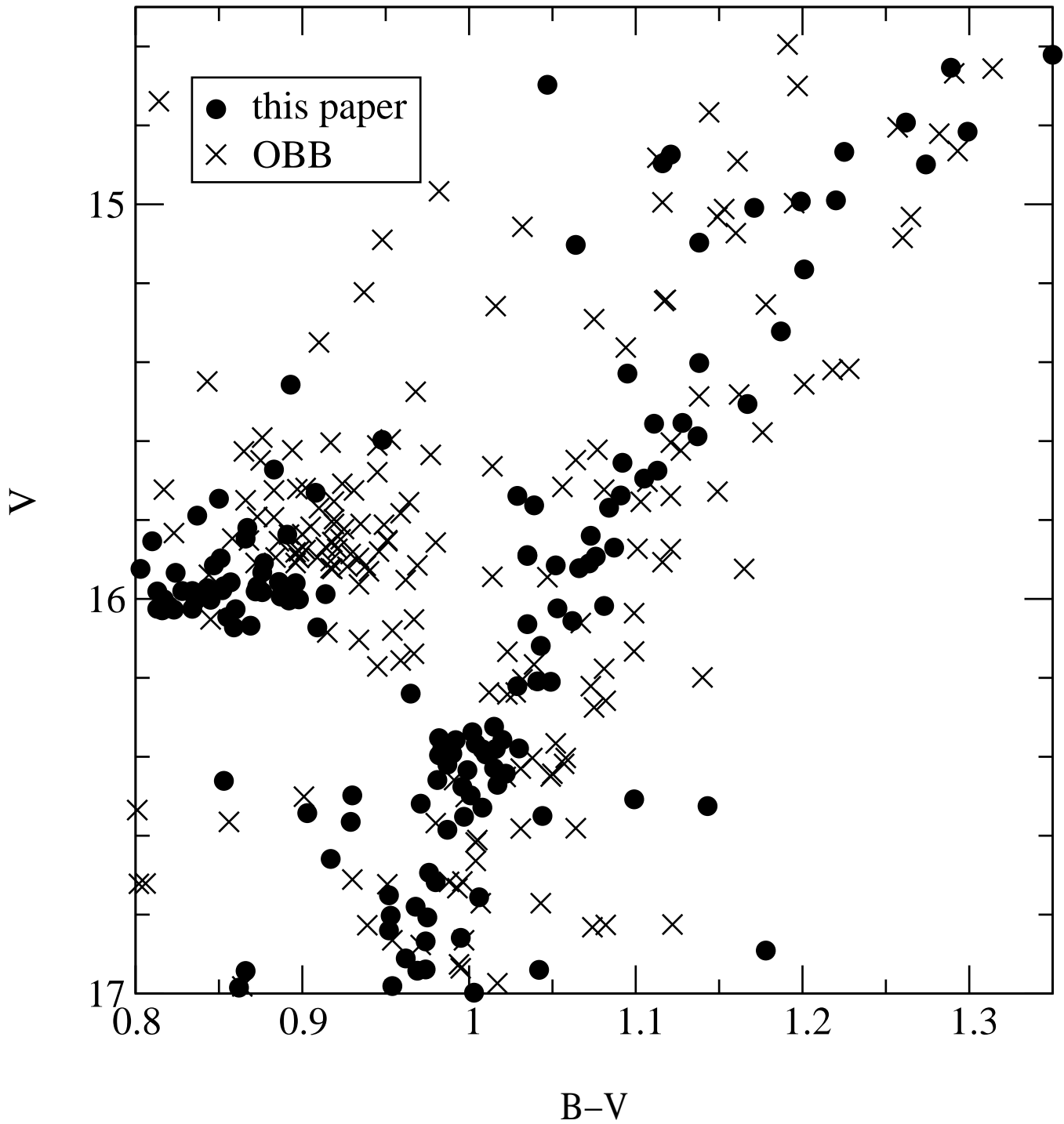}}
\caption{
A comparison of the V, \bv photometry measured from archival HST/WFPC2 
observations of NGC 6652 to that presented by OBB in the region of
the HB.
}
\label{ortbv}
\end{figure}

The ($V,    \vi$) CMD of  OBB shows   considerable scatter  around the
turn-off, so it is difficult to compare our photometry  to that of OBB
at the fainter  magnitudes.   However, it is  possible  to compare our
photometry around  the level of the  HB.  This comparison  is shown in
Figure \ref{figort}.  From this figure, it is  clear that the level of
the ZAHB is $0.03$ mag fainter in the photometry of OBB as compared to the
photometry presented in this paper.  In addition, the color of the red
giant branch (RGB) at the level of the HB as  found by OBB is 0.08 mag
bluer than the  RGB color in our  photometry.  The reason(s)  for
these differences is unclear.  The differences are considerably larger
than  our estimated errors due to  the aperture  corrections, and are
larger than  the estimated errors in the  HST $V,\vi$ calibration.  To
investigate this further, F439W  and F555W (B,V) GO 6095  observations
\citep{sosin} were  retrieved from  the  HST archive and  the
data were  reduced in  exactly the  same manner as   was done with our
V,\vi\ data.  The results  of this  comparison  are plotted  in Figure
\ref{ortbv}.  From  this, one can see that  the B--V RGBs are in
better agreement between the two  datasets than those in V--I.  
There is a slight offset in  the level of the ZAHB,
with the  OBB  photometry   being  {\sl    brighter}  than  the    HST
photometry. From the OBB \bv\ data, we obtained $<V_{\rm HB}> = 15.85$
and $(\bv)_g = 1.11$ (color of the RGB  at the level  of the HB).  From
the VI data  presented  by OBB, we find   $<V_{\rm HB}> =  15.95$  and
$(\vi)_g =  1.11$.  The disagreement between the  level of the  HB in
the  \bv\  and \vi\   datasets suggests that     the OBB data   is not
internally consistent.  In  the \bv\ and \vi\ HST  data, the level  of
the HB agrees to within 0.01 mag.  In  addition, OBB found $(\bv)_g =
(\vi)_g$.  In  general, the \vi\ color   is redder  than the
\bv\ color.  Our isochrones show  this trend, as does the observational
transformation for globular cluster giant  branches  from \bv\ to  \vi\
presented  by \citet{zinn96}. From  the  HST data, we  find $(\bv)_g =
1.06$ and $(\vi)_g = 1.22$.  The \citet{zinn96} transformation applied
to the  HST \bv\ data implies $(\vi)_g  = 1.24$ in good agreement with
the measured value.

Finally, we note that OBB determined a brighter turn-off magnitude,
$V_{\rm TO} \approx 19.2\pm 0.15$ and found $\dv \approx 3.35\pm
0.16$.  These values differ by $2.5\,\sigma$ and $2.1\,\sigma$
respectively from the values obtained with our photometry. As the
turn-off magnitude is a primary age indicator, the fainter turn-off
found in our photometry implies an older age for NGC 6652.  

\section{The Age of NGC 6652 \label{age}}
\subsection{Isochrone Fitting \label{isofit}}
\begin{figure}
\centerline{\epsfig{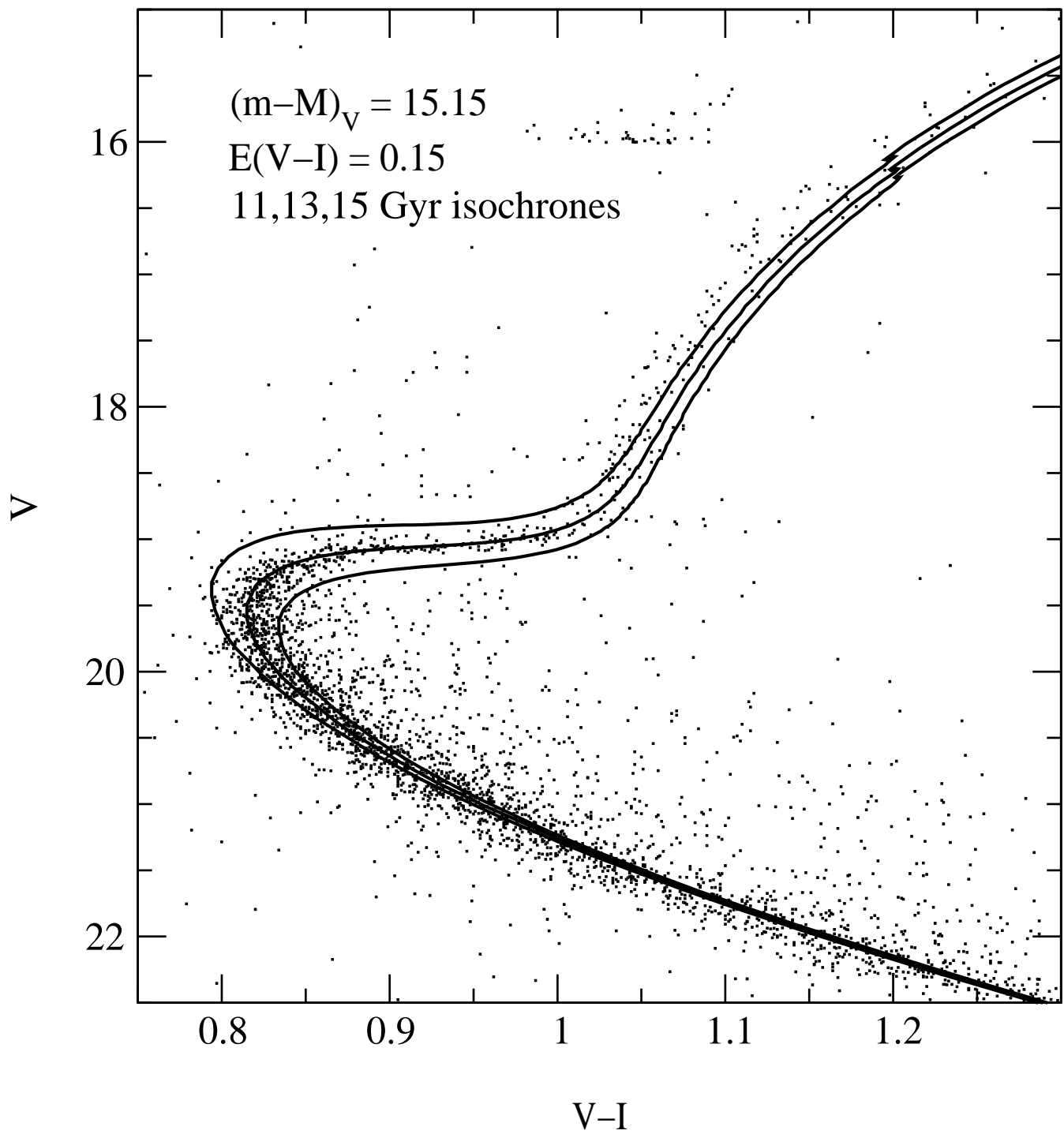}}
\caption{The fit of the $\feh = -0.85$, $[\alpha/{\rm Fe}] = +0.4$
isochrones to the data.}
\label{figfit1}
\end{figure}

A variety of methods may be used to determine the age of a globular
cluster, each with their advantages and disadvantages.  Perhaps the
simplest method is to fit an isochrone to the data by adjusting the
distance modulus and reddening (within their known uncertainties) such
that the mean locus of observed points is matched to the theoretical
isochrone.  Due to uncertainties in surface boundary
conditions, color-effective temperature transformations and the
treatment of convection, the colors of the isochrones may be in error.
This is particularly true on the RGB, where a modest change in the
choice of the mixing length can lead to a large change in the
predicted RGB colors.  For this reason, when fitting the isochrones to
the data only the main sequence and turn-off regions were used in the
fit.  Such a fit is shown in Figure \ref{figfit1} for our $\feh =
-0.85$, $[\alpha/{\rm Fe}] = +0.4$ isochrones.  The best fitting
isochrone has an age of 13 Gyr.  The best fit distance modulus ($\dmv
= 15.15$) implies $M_V(HB) = 0.81\pm 0.04$.  This is in good agreement
with the distance modulus derived using the preferred \mvrr\ relation
given by \citet{chadist} which implies $M_V(HB) = 0.73\pm 0.12$ at the
metallicity of NGC 6652.  The best fit reddening of $\evi = 0.15$
corresponds to $\ebv = 0.12$ and is in reasonable agreement with the
value obtained by OBB ($\ebv = 0.10\pm 0.02$).  The \citet{sfd98}
reddening maps based on IRAS/DIRBE dust emission yield $\ebv = 0.114$
for NGC 6652, also in good agreement.

When performing main sequence fitting, there is considerable
degeneracy between the derived distance modulus and the reddening.
For example, changing the reddening to $\evi = 0.13$, leads to a
derived distance modulus of $\dmv = 15.05$ and an age of 15 Gyr.
Assuming an uncertainty of the reddening of $\pm 0.02$ mag leads to an
uncertainty in the derived distance modulus of $\pm 0.10$ mag and in
the age of $\pm 2$ Gyr.  In fitting the isochrones to the data, the
degeneracy between the distance modulus and reddening was broken by
examining the region around the turn-off and the sub-giant branch to
obtain the best fit shown in Figure \ref{figfit1}. This allows us to
determine our preferred values for the age and distance modulus, but
it is important to realize that the error in these best fit parameters
are rather large.

The implicit assumption in this fitting procedure is that the location
of the unevolved main sequence predicted by the theoretical isochrones
is accurate.  The validity of this assumption has been discussed at
the end of \S \ref{theory}, where we conclude that at the level of
$\approx \pm 0.10$ mag the isochrones correctly predict the location
of the unevolved main sequence in relatively metal-poor stars ($-1.5
\le \feh \le 0.6$).  Combining the uncertainty in the theoretical
location of the unevolved main sequence with the uncertainty in our
fitting procedure due to errors in the reddening leads us to conclude
that the  age and distance modulus we derive from isochrone fitting
are accurate to $\pm 15\%$ and $\pm 0.15\,$mag respectively.

Fits of the isochrones with the various compositions discussed in \S
\ref{theory} were also performed.  In all cases, the fits in the
turn-off, sub-giant branch and RGB regions were considerably inferior
to the fit shown in Figure \ref{figfit1} which uses our best estimate
for the composition ($\feh = -0.85$; $[\alpha/{\rm Fe}] = +0.4$).  The
best fitting parameters for the various isochrone fits are shown in
Table \ref{tabisofit}. Given that the other compositions gave inferior
fits to the data, we prefer to take as our central values those found
from the isochrones with the best estimate for the compositions.
Taking into account the uncertainty in the reddening, along with the
variation in derived parameters shown in Table \ref{tabisofit},
isochrone fitting implies the following parameters for NGC 6652 $\dmv
= 15.15\pm 0.15$, $\evi = 0.15\pm 0.02$ and an age of $13\pm 2\,$Gyr.
Due to uncertainties in the theoretical models and the reddening of
NGC 6652, the age we derive based upon isochrone fitting has a
relatively large error.  The isochrone fitting procedure serves as a
first order test of our models (ensuring they are in reasonable
agreement with the observations) and  allows us to determine the
age of the cluster which is independent of the HB.   

\subsection{\dvbto\ Ages \label{btoage}}
A more robust determination of the cluster's age may be found using
the difference in magnitude of the point on the sub-giant branch which
is $0.05$ mag redder than the turn-off and the HB \citep[referred to
as \dvbto]{cha96}.  The magnitude of the SGB point as a function of
age is determined from the isochrones.  The theoretical value of
$M_{\rm V}({\rm HB})$ was calculated using $\mvrr = 0.23 * (\feh + 1.6)
+ 0.56$ \citep{chadist} and corrected for the fact that the HB was
only apparent redward of the RR Lyrae instability strip in NGC 6652.
This correction was determined using the theoretical HB models of
\citet{Dem00}.  From this procedure, we obtain an age of $11.7\pm
1.6\,$Gyr, where the error includes a $\pm 0.1\,$dex uncertainty in
\feh, the observational error in determining \dvbto, and the error in
our adopted \mvrr\ calibration.  This may be compared to an age of
$13\pm 2$ Gyr which was found from isochrone fitting.  This age
difference of $1.3\pm 2.6$ Gyr is well within our errors.  From the
theoretical point of view, \dvbto\ is a more robust age indicator than
isochrone fitting as it does not depend critically on the (uncertain)
colors of the theoretical models \citep{cha96}.  Thus, we prefer the
use of the \dvbto\ age indicator. Ignoring the error in the zero-point
calibration of our \mvrr\ relation gives a precise {\sl relative} age
for NGC 6652 of $11.7\pm 1.0\,$Gyr.

The above relative age may be compared to other clusters with a
similar metallicity.  For example, there exists excellent \vi\
photometry of the thick disk globular cluster 47 Tuc from \citet{47tuc}.
This cluster has a well established metallicity of 
$\feh = -0.71$ on both the \citet{zinnwest} and \citet{gratton} scales,
which is 0.14 dex more metal-rich than our preferred \feh\ for NGC
6652.  This difference in metallicity precludes the use of the
$\delta$-color technique to determine an age difference for the two clusters.

The electronic data of \citet{47tuc} was used to determine $(\vi)_{\rm
TO} = 0.714\pm 0.004$, $V_{\rm SGB} = 17.25\pm 0.03$, $<V_{\rm HB}> =
14.07\pm 0.04$, $\dvbto = 3.18\pm 0.05$. These values were determined
in exactly the same manner as for NGC 6652.  The value of \dvbto for
47 Tuc is very similar to that of NGC 6652.  Formally, $\dvbto ({\rm
N6652 - 47Tuc}) = -0.05\pm 0.07$.  Assuming that 47 Tuc has
$[\alpha/{\rm Fe}] = +0.40$ and a helium abundance which is given by
eqn. (\ref{yeqn}) then a precise relative age of $12.2\pm 0.7$ Gyr is
obtained for 47 Tuc using the \dvbto\ method.  This is virtually
identical to the age derived for NGC 6652.  To the precision of the
data, the two clusters have the same age to within $\pm 1.2\,$Gyr.

\citet{Bro92} found an enhancement in $\alpha$-capture elements of
$[\alpha/{\rm Fe}] = 0.22$ in 47 Tuc.  Using this value of
$[\alpha/{\rm Fe}]$ and using the same assumptions as this paper,
\citet{wilson} found a \dvbto\ age of $12.9\pm 0.7\,$Gyr for 47 Tuc
implying that 47 Tuc is $1.2\pm 1.2\,$Gyr older than NGC 6652.  If 47
Tuc has a solar helium abundance, the absolute magnitude of the HB
would be different than what has been assumed, implying that the
derived age would be altered.  \citet{sw98} found that increasing the
helium abundance by 0.019 to (their) solar value decreased the
estimated age of 47 Tuc by 1.1 Gyr. This difference in helium
abundance between our solar value and that implied by
eqn. (\ref{yeqn}) is 0.0177, and so if 47 Tuc has a solar helium
abundance the \dvbto\ age we determined should be decreased by 1.0
Gyr.  This would imply that the NGC 6652 is $0.9\pm 1.2\,$Gyr older
than 47 Tuc. Thus, allowing for differences in the $[\alpha/{\rm Fe}]$
or in the helium abundance between the two clusters does not
change our conclusion that, to the precision of the data, NGC 6652 and
47 Tuc are the same age.  This lack of age difference between 47 Tuc
and NGC 6652 implies that the formation of the thick disk overlapped
with the formation of the inner halo.

\begin{figure}
\centerline{\epsfig{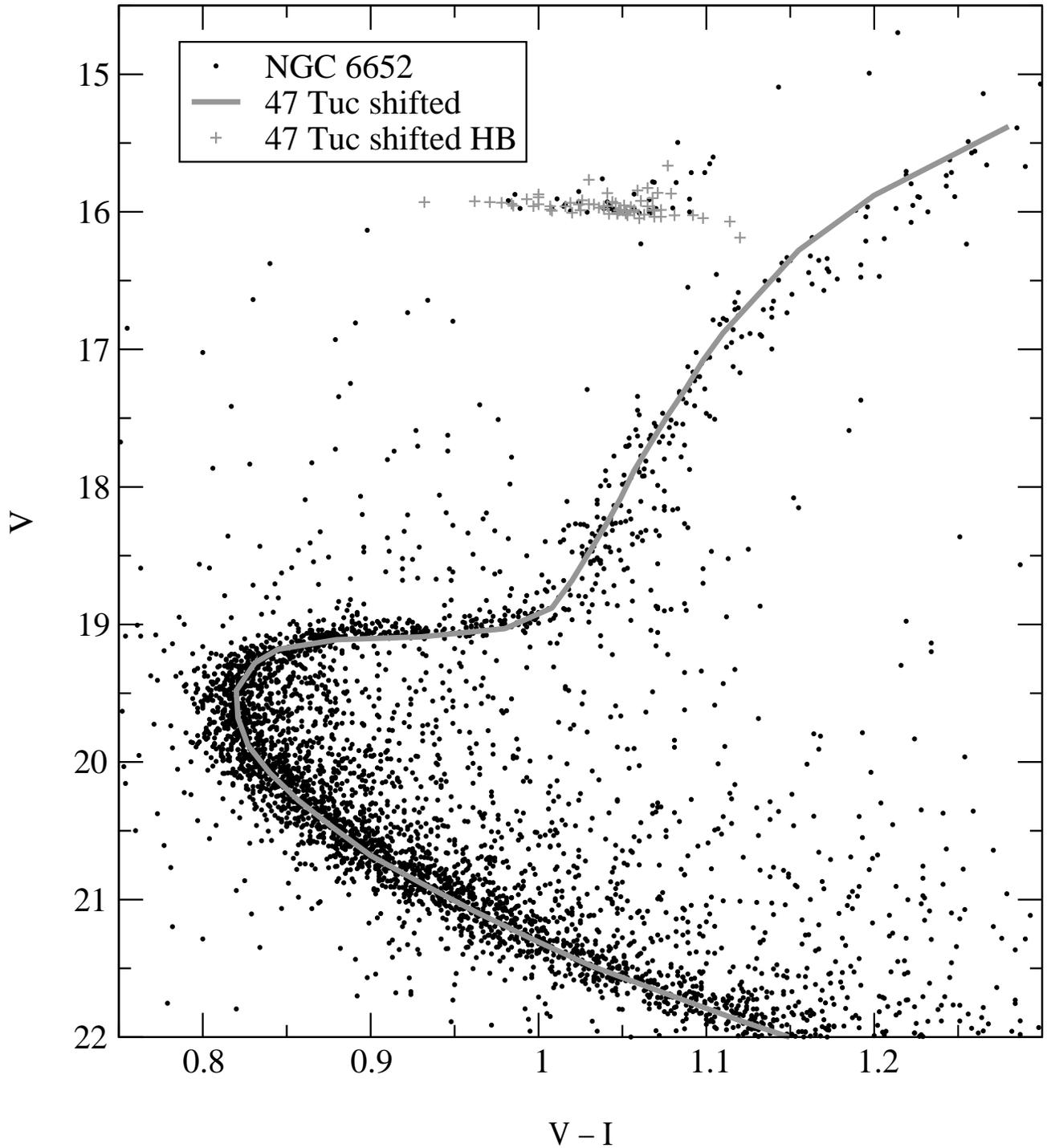}}
\caption{A comparison between our data for NGC 6652 and the
\citet{47tuc} data for 47 Tuc.  To facilitate the comparison, a
fiducial for the 47 Tuc main sequence and RGB was determined and is
shown on the graph.  The 47 Tuc data was shifted by $\Delta V =
+1.88$\, mag and $\Delta \vi = +0.105$ mag.  These values were
determined by requiring that the turnoff colors and the luminosities
of the observed ZAHBs be coincident between the two clusters.}
\label{figcompare}
\end{figure}

At the suggestions of the referee, we have tested the results of our
\dvbto\ analysis by overlaying the CMD of 47 Tuc with that of NGC
6652.  The 47 Tuc data was shifted so that the turnoff colors and the
luminosities of the observed ZAHBs are made coincident.  The results
of this comparison are shown in Figure \ref{figcompare}.  A careful
inspection of this figure reveals that when the ZAHBs of the 2
clusters are made coincident, the 47 Tuc subgiant branch is slightly
fainter (by $0.06\pm 0.03$ mag) than the subgiant branch of NGC 6652.
Thus, \dvbto\ for 47 Tuc is slightly larger than \dvbto\ in NGC 6652.
This is exactly what was determined by the \dvbto\ analysis presented
in the previous paragraph.

The age of NGC 6652 may also be compared to NGC 6171 (M107).  M107 is
also considered a member of the halo \citep{zinn85}.  From
their Ca II triplet analysis, \citet{rut} found $\feh = -1.09$ on the
\citet{zinnwest} scale and $\feh = -0.95$ on the \citet{gratton}
scale.  This implies that M107 is $\simeq 0.1$ dex more metal-poor
than NGC 6652.  A V, \bv\ CMD for M107 has been presented by
\citet{m107}.  As no V, \vi\ data is publicly available for this
cluster, we cannot do a differential age comparison using the
$\delta$-color technique, and instead analyze the age difference
using \dvbto.  The electronic data of \citet{m107} was carefully
inspected, and the following points measured: $(\bv)_{\rm TO} =
0.870\pm 0.01$, $V_{\rm SGB} = 18.86\pm 0.04$, $<V_{\rm HB}> =
15.59\pm 0.03$, $\dvbto = 3.27\pm 0.05$.  This value of \dvbto\ is
$0.14\pm 0.07$ mag larger than that found for NGC 6652 ($\dvbto =
3.13\pm 0.05$).  Using our isochrones, an \dvbto\ age for M107 of
$14.0\pm 1.1\,$Gyr is found, which is $2.7\pm 1.5\,$Gyr older than the
\dvbto\ age of NGC 6652.  This suggests that M107 may be older than
NGC 6652.  However, this result is only significant at the
$1.8\,\sigma$ level. A more definitive study (using \vi\ photometry
for M107) is needed to determine if NGC 6652 is younger than M107.

To investigate this issue further, we have also determined the age of
NGC 1851, a somewhat more metal-poor cluster.  \citet{rut} found $\feh
= -1.23$ on the \citet{zinnwest} scale and $\feh = -1.03$ on the
\citet{gratton} scale, implying that NGC 1851 is $\simeq 0.2\,$dex
more metal-poor than NGC 6652.  \citet{walker} obtained BVI CCD data
for this cluster.  Using the $V$, \vi\ data, we determined $(\vi)_{\rm
TO} = 0.581\pm 0.01$, $V_{\rm SGB} = 19.11\pm 0.04$, $<V_{\rm HB}> =
16.12\pm 0.04$, $\dvbto = 2.99\pm 0.06$.  This implies an age of
$10.4\pm 1.0$ Gyr, adopting $\feh = -1.10$.  Thus, NGC
1851 is $0.9\pm 1.4\,$Gyr {\sl younger} than NGC 6652, and $3.6\pm
1.5\,$Gyr younger than M107.  \citet{sw98} and \citet{rosenberg} have
also concluded that NGC 1851 is somewhat younger than M107 ($2.1\pm 1.4\,$Gyr
and $2.9\pm 1.3$\, Gyr respectively,  using
different age diagnostics).  In summary, there is a suggestion that
NGC 6652 is somewhat younger ($2.7\pm 1.5\,$Gyr) than the halo cluster
M107, while it appears to be slightly older ($0.9\pm 1.4\,$Gyr) than
NGC 1851.  

\section{Summary \label{summ}}
HST ($V,\, I$) photometry has been obtained for the inner halo
globular cluster NGC 6652.  This photometry includes a well populated
HB and extends well below the main sequence turn-off.  From these data,
we determined  $V_{\rm ZAHB} = 16.00\pm 0.03$,
$<V_{\rm HB}> = 15.96\pm 0.04$, $(\vi)_{\rm TO} = 0.818\pm 0.004$,
$V_{\rm TO} = 19.55\pm 0.07$,  $V_{\rm SGB} = 19.09\pm 0.03$,
$\dv =  3.59\pm 0.08$, and $\dvbto = 3.13\pm 0.05$.   This turn-off
magnitude is $0.35\pm 0.17$ magnitudes fainter than that found in the
previous study of NGC 6652 by OBB.  As a consequence, the age derived
from our photometry is substantially older than age determinations
based upon the OBB photometry.  The OBB photometry shows considerable
scatter around the turn-off, while the data presented here clearly
delineates the main sequence turn-off region.

New stellar evolution models and isochrones were calculated for $\feh =
-1.20,\, -1.00,\, -0.85$ and $-0.70$ in order to determine the age of
NGC 6652 and other globular clusters with similar metallicities.
Estimates for the metallicity of NGC 6652 vary from $\feh = -0.96$ on
the \citet{zinnwest} scale to $\feh = -0.85$ on the \citet{gratton}
scale.  Our best fitting isochrones have $\feh = -0.85$ and imply
$\dmv = 15.15\pm 0.10 $ and $\evi = 0.15\pm 0.02$.  With $\rgc \simeq
2.0\,$kpc, NGC 6652 is the globular cluster closest to the Galactic
center for which a precise relative age comparison can be made to
other globular clusters which are relatively metal-poor.  From the HST
data, NGC 6652 has an  age of $11.7\pm 1.6\,$Gyr (using  \dvbto\ as
the age indicator).  Using data
from the literature, \dvbto\ ages of $12.2\pm 0.7$ Gyr for 47 Tuc (thick
disk cluster with $\feh = -0.71$), $14.0\pm 1.1\,$Gyr for M107 (halo
cluster with $\feh = -0.95$) and $10.4\pm 1.0$ Gyr for NGC 1851 (halo
cluster with $\feh \simeq -1.1$) were determined.  These precise
relative ages demonstrate that the halo clusters NGC 6652 and NGC 1851
are the same age as the thick disk cluster 47 Tuc.  There is some
evidence that M107 is somewhat older, but the difference in age
between M107 and NGC 6652 is only significant at the $1.5\,\sigma$
level.  A more definitive study (using \vi\ photometry for M107) is
needed to determine if NGC 6652 is younger than M107.

\acknowledgements 
We would like to thank the anonymous referee whose
suggestions led to an improved paper.  This research was supported by
NASA through grant number GO-06517 from the Space Telescope Science
Institute, which is operated by AURA, Inc., under NASA contract NAS
5-26555.

\clearpage

\begin{deluxetable}{lrrrc}
\tablecaption{Photometric Data \label{tabphot}}
\tablewidth{0pt}
\tablehead{
\colhead{Chip} &
\colhead{X (pixel)} & 
\colhead{Y (pixel)} &
\colhead{V (mag)}&
\colhead{\vi (mag)}
}
\startdata
 PC1&  420.61&  462.47&  14.698&   1.214\\  
 PC1&  665.83&  481.91&  14.852&   1.434\\
 PC1&  731.93&  164.93&  14.929&   1.362\\
 PC1&  415.13&  514.46&  14.969&   1.347\\
 PC1&  290.54&  521.16&  15.071&   1.299\\
 PC1&  277.56&  709.88&  15.174&   1.322\\  
 PC1&  250.74&  585.43&  15.413&   1.303\\  
 PC1&  382.62&  258.89&  15.489&   1.256\\  
 PC1&  487.84&  607.94&  15.496&   1.083\\  
 PC1&  390.07&  525.26&  15.560&   1.260\\  
 PC1&  639.05&  254.91&  15.623&   1.245\\  
 PC1&  269.85&  435.04&  15.651&   1.102\\  
 PC1&  446.00&  487.11&  15.659&   1.267\\  
 \enddata
\end{deluxetable}

\begin{deluxetable}{ccccccc}
\tablecaption{Isochrone Fit Parameters \label{tabisofit}}
\tablewidth{0pt}
\tablehead{
\colhead{\feh} & 
\colhead{$[\alpha/{\rm Fe}]$} &
\colhead{$Y$}&
\colhead{\dmv}&
\colhead{\evi}&
\colhead{Age (Gyr)}&
\colhead{Note}
}
\startdata
 $-0.85$ & 0.40 & 0.2420 & 15.15 & 0.15 & 13 & best fit\\
 $-0.85$ & 0.40 & 0.2420 & 15.05 & 0.13 & 15 & good fit\\
 $-0.85$ & 0.00 & 0.2379 & 15.09 & 0.17 & 14 & poor fit to RGB\\
 $-1.00$ & 0.40 & 0.2395 & 15.13 & 0.16 & 14 & poor fit to SGB\\
 $-0.70$ & 0.40 & 0.2453 & 15.12 & 0.12 & 13 & poor fit to RGB\\
 $-0.70$ & 0.40 & 0.2630 & 15.12 & 0.12 & 13 & poor fit to RGB\\
 \enddata 
\end{deluxetable}

\clearpage

\clearpage

\clearpage

\clearpage

\clearpage

\clearpage

\clearpage

\end{document}